\documentclass[hyper]{JHEP} % 10pt is ignored!
\usepackage{epsfig}
\usepackage{amsfonts}
\newcommand\fverb{\setbox\pippobox=\hbox\bgroup\verb}
\newcommand\fverbdo{\egroup\medskip\noindent%
            \fbox{\unhbox\pippobox}\ }

\newcommand\fverbit{\egroup\item[\fbox{\unhbox\pippobox}]}
\newbox\pippobox
\title{Pulsating Strings With Angular Momenta}
\preprint{1306.0457[hep-th]}
\author{Pabitra M. Pradhan\\
Department of Physics \&  Meteorology, \\
Indian Institute of Technology Kharagpur,\\
Kharagpur-721 302, INDIA \\
E-mail: \email{ppabitra@phy.iitkgp.ernet.in}}
\author{Kamal L Panigrahi\\
Department of Physics \& Meteorology,\\
Indian Institute of Technology Kharagpur,\\
Kharagpur-721302, INDIA,\\
and\\
The Abdus Salam International Centre for Theoretical Physics,\\
Strada Costiera 11, Trieste, ITALY\\
E-mail:\email{panigrahi@phy.iitkgp.ernet.in}} \abstract{We derive
the energy of pulsating string, as function of oscillation number
and angular momenta, which oscillates in $AdS_3$ with an extra
angular momentum along $S^1$. We find similar solutions for the
strings oscillating in $S^3$ in addition to extra angular
momentum. Further we generalize the result of the oscillating
strings in Anti de-Sitter space in the presence of both spin and
angular momentum in $AdS^5 \times S^1$.} 

\begin{document}
\section{Introduction}
Proving the AdS/CFT correspondence
\cite{Maldacena:1997re}-\cite{Witten:1998qj} or more generally the
gauge-gravity duality has been a  major area of research in string
theory in the last many years. This duality maps the anomalous
dimensions of gauge invariant operators in the gauge theory to the
energy spectrum of the string theory states. It has been observed
that the exact state-operator mapping is extremely difficult,
because of the infinite number of stringy states in the string
theory side. Hence various limits have been considered on both
sides of the duality, so that the conjecture can be checked to
some accuracy. The semiclassical strings have played a key role in
exploring various aspects of the correspondence
\cite{Bena:2003wd}-\cite{Beisert:2006ez}. This shows that the
semiclassical results are reliable enough to use for the duality
and compare with the dimensions of the operators in the Super
Yang-Mills (SYM) side. The matrix of anomalous dimensions can be
mapped to an integrable Bethe spin chain \cite{Minahan:2002ve},
and it has empowered our understanding of the duality conjecture.
The semiclassical calculations in the string theory side has shown
that the multi-spin rotating and pulsating string solutions beyond
their BPS limit with large charges are in perfect agreement with
the ones calculated in dual gauge theory. The agreement beyond the
BPS limit relies on the fact that on the string theory side the
quantum corrections of strings are suppressed by the large quantum
number, while in the field theory side the anomalous dimension
matrices of the dual composite operators are related to the
Hamiltonian of integrable spin chain. Among the classes of the
semiclassical string solutions, the pulsating string solutions
have better stability than the non-pulsating ones
\cite{Khan:2005fc}. However unlike rotating strings, pulsating
string are less explored. Pulsating string solution was first
introduced in \cite{Minahan:2002rc} and further generalized in
\cite{Engquist:2003rn}-\cite{Smedback:1998yn}. They have also been
studied in $AdS_5 \times S^5$
\cite{Gubser:2002tv}-\cite{Panigrahi:2012in}, $AdS_4 \times CP^3$
\cite{Chen:2008qq}-\cite{Dimov:2009rd} and other backgrounds
\cite{Bobev:2004id}-\cite{Giardino:2011jy}. Pulsating strings
concept was first introduced in \cite{Gubser:2002tv} where they
were expected to correspond to certain highly excited sigma model
operators. In \cite{Minahan:2002rc} and \cite{Beccaria:2010zn},
pulsating string solutions in $AdS_5$ and $S^5$ have been worked
out separately where as in \cite{Park:2005kt}, simultaneously
rotating and oscillating strings in $AdS_5$ have been derived.
Strings spinning in $S^5$ where as pulsating in $AdS$ and spinning
in $AdS$ where as pulsating in $S^5$ have been described in
\cite{Khan:2003sm}. Recently pulsating string solution in the less
supersymmetric Lunin-Maldacena background has been studied in
\cite{Giardino:2011jy} and dispersion relation for the string
oscillating in $S^5$ with a constant $\rho$ value in $AdS_5$ has
been found out in \cite{Panigrahi:2012in}. As oscillation number
is adiabatic invariant, the relation between energy and
oscillation number is presented as the solutions to characterize
the string dynamics. So here we wish to study a few oscillating
string solutions with an extra angular momentum and see how the
energy and oscillation number relation is affected when the string
is oscillating in $S$ and $AdS$ separately.

The rest of the paper is organized as follows. In the section-2,
we show the relationship of the  oscillation number with an
angular momentum for the string which is pulsating in the radial
direction of $AdS$ with an angular momentum along $S^1$. In the
section-3, we present a class of circular string solution which is
oscillating in $S^3$ and at the same time has an extra angular
momentum. Section-4 is devoted to the rotating string which is
pulsating in the $AdS$ with two spin along mutually perpendicular
directions of $AdS_5$ and have an angular momentum along $S^1$.
Finally, in the section-5, we conclude with some remarks.
\section{Pulsating string with angular momentum
in $AdS_3 \times S^1$} In this section we study a semiclassical
quantization of a string which expands and contracts in AdS$_3$
and at the same time rotates along sphere. We start with the full
metric for $AdS_5 \times S^5$ background
\begin{eqnarray}
ds^2 &=& -\cosh^2\rho dt^2 + d\rho^2 + \sinh^2\rho (d\phi^2 +
\cos^2\phi d\phi^2_1 + \sin^2 \phi d\phi^2_2) + d\psi^2 \cr & \cr
&& + \sin^2\psi d\theta^2 +\cos^2\psi d\xi^2 +\sin^2\psi
\cos^2\theta d\xi^2_1 + \sin^2\psi \sin^2\theta d\xi^2_2 \ .
\label{A}
\end{eqnarray}
By appropriate substitution of the co-ordinates in the above
metric ({\ref{A}), we get the following metric for studying the
string which oscillates in the radial ($\rho$) direction of $AdS$
space and at the same time has an angular momentum along $S^1$:
\begin{eqnarray} ds^2 &=& -\cosh^2\rho dt^2 + d\rho^2 +
\sinh^2\rho d\phi^2 + d\psi^2, \label{1}
\end{eqnarray}
where $\rho \in [0,\infty],\phi \in [0,2\pi]$ and $ \psi$ specifies
$S^1$ direction. The Polyakov action for the fundamental
string in the above background is given as
\begin{equation}
I = \frac{\sqrt\lambda}{4\pi}\int d\tau d\sigma\Big[-\cosh^2\rho
(\dot{t}^2 - {t^{\prime}}^2) + \dot{\rho}^2 - {\rho^{\prime}}^2 +
\sinh^2\rho(\dot{\phi}^2 - {\phi^{\prime}}^2) +  \dot{\psi}^2 -
{\psi^{\prime}}^2 \Big], \label{2}
\end{equation}
where $\lambda$ is the 't Hooft coupling, `dots' and `primes'
denote the derivatives with respect to world sheet time and space
coordinates respectively. We take the following ansatz for
studying the pulsating string solution
\begin{eqnarray}
&&t = t(\tau), ~~~~ \rho = \rho(\tau), ~~~~ \phi = m \sigma,~~~~
\psi = \psi(\tau). \label{3}
\end{eqnarray}
The equations of motion for $t$ and $\rho$ are given by
\begin{eqnarray}
&&2\cosh\rho \sinh\rho \dot{\rho}\dot t + \cosh^2\rho \ddot t =
0,\cr & \cr &&\ddot\rho = -\sinh\rho \cosh\rho
(m^2+\dot{t}^2).\label{4}
\end{eqnarray}
The Virasoro constraints gives us the following
\begin{equation}
\dot\rho^2 + \dot\psi^2 + m^2\sinh^2\rho - \cosh^2\rho \dot t^2 =
0. \label{5}
\end{equation}
The conserved quantities are as follows
\begin{equation}
\mathcal E = \cosh^2\rho \dot t, ~~~~~~~~~~~ \mathcal J =
\dot\psi, \label{6}
\end{equation}
where $\mathcal E = \frac{E}{\sqrt\lambda}$ and $\mathcal J =
\frac{J}{\sqrt\lambda}$. Now putting the above equation (\ref{6})
in the equation(\ref{5}), we get
\begin{equation}
\dot\rho^2 = \frac{\mathcal E^2}{\cosh^2\rho} - m^2\sinh^2\rho
-\mathcal J^2. \label{7}
\end{equation}
Some comments are in order. In the above equation, we can see that
$\dot\rho^2$ goes from $\mathcal E^2 - \mathcal J^2$ to infinity
as $\rho$ goes from 0 to $\infty$. This  means that the coordinate
$\rho$ oscillates between a minimal value $(0)$ and a maximal
value ($\rho_{max}$). Now the oscillation number can be written as
\begin{equation}
N = \frac{\sqrt\lambda}{2\pi} \oint d\rho \dot\rho = \frac{1}{\pi}
\int_{0}^{\rho_{max}} d\rho \sqrt{\frac{\mathcal E^2}{\cosh^2\rho}
- m^2\sinh^2\rho -\mathcal J^2 }.\label{8}
\end{equation}
Putting $\sinh\rho = x$ in the above equation (\ref{8}), we get
\begin{equation}
\mathcal N = \frac{1}{\pi} \int_{0}^{\sqrt R} \frac{dx}{1+x^2}
\sqrt{\mathcal E^2 -m^2 x^2(1+x^2) - \mathcal J^2(1+x^2)},
\label{9}
\end{equation}
where $\mathcal N \sqrt\lambda = N$ and $R= \frac{-(m^2+ \mathcal
J^2)+ \sqrt{(m^2+\mathcal J^2)^2 + 4 m^2 (\mathcal E^2 -\mathcal
J^2) }}{2 m^2}$. Taking the partial derivative of the above
equation with respect to $m$ we get
\begin{equation}
\frac{\partial \mathcal N}{\partial m} = -\frac{m}{\pi}
\int_{0}^{\sqrt R}\frac{x^2}{\sqrt{\mathcal E^2 -m^2 x^2(1+x^2) -
\mathcal J^2(1+x^2)}}dx. \label {10}
\end{equation}
The above can be written in terms of the standard elliptical
integrals
\begin{equation}
\frac{\partial \mathcal N}{\partial m} = \frac{1}{\sqrt 2\pi m}
\sqrt a_+ \left[\mathbb {K}\Big(\frac{a_-}{a_+}\Big) - \mathbb
{E}\Big(\frac{a_-}{a_+}\Big)\right], \label{11}
\end{equation}
where $a_\pm = (m^2+\mathcal J^2) \pm \sqrt{(m^2+\mathcal J^2)^2 +
4 m^2(\mathcal E^2- \mathcal J^2)}$ and $\mathbb {K}$ and $\mathbb
{E}$ are complete elliptical integral of first and second kind
respectively. Now expanding the above equation for small
oscillation number with small $\mathcal E$ and $\mathcal J$
\begin{eqnarray}
\frac{\partial \mathcal N}{\partial m} &=& \left[\frac{\mathcal
J^2} {4m^2} + \frac{3\mathcal J^4}{32 m^4} + \mathcal O[\mathcal
J^6]\right] + \left[-\frac{1}{4 m^2} - \frac{9 \mathcal J^2}
{16m^4} - \frac{225\mathcal J^4}{256 m^6} + \mathcal O[\mathcal
J^6]\right] \mathcal E^2 \cr && \cr &+& \left[\frac{15}{32
m^4}+\frac{525\mathcal J^2} {256m^6} + \frac{11025\mathcal
J^4}{2048 m^8} + \mathcal O[\mathcal J^6]\right] \mathcal E^4 +
\mathcal O[\mathcal E^6]. \label{12}
\end{eqnarray}
Integrating the above equation (\ref{12}) with respect to $m$ and
reversing the series, we get
\begin{equation}
\mathcal E = 2 \sqrt{m \mathcal M}~~ K_1(\mathcal J)\left[ 1 +
K_2(\mathcal J) \frac{5\mathcal M}{4m}+ \mathcal O[\mathcal
M^2]\right], \label{13}
\end{equation}
where
\begin{eqnarray}
 ~~~\mathcal M &=& \mathcal N + \frac{\mathcal J^2}{4
m}+\frac{\mathcal J^4}{32 m^3}+ \mathcal O[\mathcal J^6] \cr &&
\cr K_1(\mathcal J) &=& \left[1 + \frac{3\mathcal J^2}{4
m^2}+\frac{45\mathcal J^4}{64 m^4}+ \mathcal O[\mathcal
J^6]\right]^{-1/2} \cr && \cr {\rm and}~~~ K_2(\mathcal J) &=& \left[1 +
\frac{21\mathcal J^2}{8 m^2}+\frac{315\mathcal J^4}{64 m^4}+
\mathcal O[\mathcal J^6]\right] K^4_1(\mathcal J). \label{14}
\end{eqnarray}
The equation (\ref{13}) represents the classical energy for the
short strings which are oscillating near the center of AdS$_3$
with an angular momentum in S$^1$. With $\mathcal J \rightarrow
0$, $\mathcal M \rightarrow \mathcal N$ and $K_{1,2} \rightarrow
1$. This gives us the energy for the strings oscillating in one
plane for small energy limit as in the \cite{Park:2005kt}. Now
expanding equation (\ref{11}), for large $\mathcal E$ but small
$\mathcal J$, we get
\begin{eqnarray}
\frac{\partial \mathcal N}{\partial m} &=& k_1 m^{-1/2}\mathcal
E^{1/2} + k_2 m^{1/2} \mathcal E^{-1/2} \Big(1+ \frac{\mathcal
J^2}{m^2} \Big) \cr && \cr &+& k_3 m^{3/2} \mathcal
E^{-3/2}\Big(1+ \frac{k_4\mathcal J^2}{k_3m^2} + \frac{\mathcal
J^4}{m^4} \Big) + \mathcal O[\mathcal E^{-5/2}], \label{15}
\end{eqnarray}
where $~~~k_1 = \frac{1}{\pi}\Big(\sqrt\pi\frac{\lceil
[\frac{5}{4}]}{\lceil [\frac{3}{4}]}-\mathbb {E}(-1)\Big) =
-0.19069$, \\
$~~~~~~~~~~~k_2 = -\frac{\mathbb {E}(-1)}{4\pi} +
\frac{1}{8\sqrt\pi} \frac{\lceil [\frac{3}{4}]}{\lceil
[\frac{5}{4}]} + \frac{1}{4\sqrt\pi} \frac{\lceil
[\frac{5}{4}]}{\lceil [\frac{3}{4}]} + \frac{1}{8}
~_2F_1(\frac{3}{2}, \frac{3}{2},2,-1)
= 0.104328$, \\
 $~~~~~~~~~k_3 =-\frac{\mathbb {E}(-1)}{32\pi} - \frac{1}{32\sqrt\pi}
\frac{\lceil [\frac{3}{4}]}{\lceil [\frac{5}{4}]} +
\frac{1}{32\sqrt\pi} \frac{\lceil [\frac{5}{4}]}{\lceil
[\frac{3}{4}]} - \frac{1}{32} ~_2F_1(\frac{3}{2},
\frac{3}{2},2,-1) \\ ~~~~~~~~~~~~~~~~~~~+ \frac{3}{128}
~_2F_1(\frac{3}{2}, \frac{5}{2},3,-1) + \frac{9}{128}
~_2F_1(\frac{5}{2}, \frac{5}{2},3,-1) = -0.0178772$,\\and $k_4 =
0.0119181$. Integrating the above equation (\ref{15}), we get
\begin{eqnarray}
\mathcal N = \mathcal N_0 &+& 2k_1 m^{1/2}\mathcal E^{1/2} +
\frac{2}{3} k_2 m^{3/2} \mathcal E^{-1/2} \Big(1 - 3
\frac{\mathcal J^2}{m^2}\Big) \cr && \cr &+& \frac{2}{5} k_3
m^{5/2} \mathcal E^{-3/2}\Big(1+ 5 \frac{k_4\mathcal J^2}{k_3 m^2}
- \frac{5}{3} \frac{\mathcal J^4}{m^4}\Big) + \mathcal O[\mathcal
E^{-5/2}]. \label{16}
\end{eqnarray}
The integration constant $\mathcal N_0$ can be determined from the
integral (\ref{9}) for $m = 0$
\begin{equation}
\mathcal N_0 = \frac{1}{\pi}
\int_{0}^{r}\frac{dx}{1+x^2}\sqrt{\mathcal E^2-\mathcal J^2
-\mathcal J^2 x^2}, \label{17}
\end{equation}
where $r = \sqrt{\frac{\mathcal E^2-\mathcal J^2}{\mathcal J^2}}$.
Changing the variable we get
\begin{equation}
\mathcal N_0 = \mathcal J \frac{r^2}{\pi} \int_{0}^{1}
\frac{\sqrt{1-x^2}}{1+r^2x^2} dx = \frac{1}{2} (\mathcal
E-\mathcal J). \label{18}
\end{equation}
Now
\begin{equation}
\mathcal N = \frac{1}{2} (\mathcal E-\mathcal J) + 2k_1
m^{1/2}\mathcal E^{1/2} + \frac{2}{3} k_2 m^{3/2} \mathcal
E^{-1/2} \Big(1 - 3 \frac{\mathcal J^2}{m^2}\Big) + \mathcal
O[\mathcal E^{-3/2}]. \label{19}
\end{equation}
Reversing the series we get
\begin{equation}
\mathcal E = 2L +  a_1 m^{1/2} L^{1/2} + a_2 m  - A(\mathcal J)
~a_3~m^{3/2} L^{-1/2} + \mathcal O [L^{-3/2}], \label{20}
\end{equation}
where ~~ $L = \mathcal N + \frac{\mathcal J}{2},~~~~~~~~~~~~
A(\mathcal J) = 1 - 5 \frac{\mathcal J^2}{m^2},\\
a_1 = 1.07871,~~~~~ a_2=0.290901 ~~$ and $~~ a_3 = 0.0591372 $. In
the above (\ref{20}) equation, from $\mathcal O[L^{-1/2}]$ term
onwards the coefficients will depend on angular momentum $\mathcal
J$ raised to the even exponent explicitly with $L$. The $\mathcal
J$ dependent terms will run up to the next even exponent of
$\mathcal J$ i.e. the $\mathcal J$ dependent term of $\mathcal
O[L^{-3/2}]$ will run up to $\frac{\mathcal J^4}{m^4}$, $\mathcal
O[L^{-5/2}]$ up to $\frac{\mathcal J^6}{m^6}$ and so on. The above
equation (\ref{20}) is the energy expression for the long strings
with small $\mathcal J$ in $AdS_3 \times S^2$. This is the same
expression as in \cite{Minahan:2002rc} with $\mathcal J = 0$.
Expanding (\ref{11}) for short strings with large $\mathcal J$ we
get
\begin{equation}
\mathcal E = 2\mathcal N + \mathcal J - \frac{m^2}{4} \mathcal
J^{-1} - \frac{3m^4}{64} \mathcal J^{-3} - \mathcal O[\mathcal
J^{-5}]. \label{22}
\end{equation}
\section{Pulsating string in $\mathbb {R} \times S^3$}
Here we wish to study a class of string solutions which is
pulsating in $S^3$ as well as with an extra angular momentum. The
%We use the following So we substitute $\rho = \xi_1
%= \xi_2 = 0$ in the equation (\ref{A})and get the relevant
background metric is
\begin{equation}
ds^2 = -dt^2 + d\psi^2 + \sin^2\psi d\theta^2 + \cos^2\psi d\xi^2.
\label{23}
\end{equation}
The Polyakov action for the string in the above background is
given by
\begin{equation}
I = \frac{\sqrt\lambda}{4\pi}\int d\tau d\sigma\Big[-(\dot t^2 -
{t^\prime}^2) + \dot\psi^2-{\psi^\prime}^2 +
\sin^2\psi(\dot\theta^2 - {\theta^\prime}^2) +\cos^2\psi
(\dot\xi^2 - {\xi^\prime}^2) \Big]. \label{24}
\end{equation}
We chose the following ansatz for studying the pulsating string
with an extra angular momentum in the $\mathbb{R} \times S^3$
space
\begin{equation} t=
t(\tau), ~~~~~~ \psi =\psi(\tau), ~~~~~~ \theta =m \sigma, ~~~~~~
\xi = \xi(\tau).\label{25}
\end{equation}
The equations of motion for $\psi$ and $\xi$ are given by
\begin{eqnarray}
\ddot\psi &=& -(m^2 + \dot\xi^2)\sin\psi \cos\psi, \cr && \cr
\ddot\xi &=& 2 \dot\psi \dot\xi \tan\psi. \label{26}
\end{eqnarray}
On the other hand, the Virasoro constraints give us
\begin{equation}
\dot\psi^2 - \dot t^2 + m^2 \sin^2\psi + \cos^2\psi \dot\xi^2 = 0.
\label{27}
\end{equation}
The conserved charges are given as
\begin{equation}
\mathcal E = \dot t, ~~~~~~~~~~~ \mathcal J = \cos^2\psi \dot\xi.
\label{28}
\end{equation}
Putting above equation (\ref{28}) in the equation (\ref{27}), we
get
\begin{equation}
\dot\psi^2 = \mathcal E^2 -m^2 \sin^2\psi - \frac{\mathcal
J^2}{\cos^2\psi}.\label{29}
\end{equation}
Now the oscillation number is written as
\begin{equation}
\mathcal N = \frac{1}{2\pi} \oint d\psi \sqrt{\mathcal E^2 -
m^2\sin^2\psi -\frac{\mathcal J^2}{\cos^2\psi}}.\label{30}
\end{equation}
Putting $\sin\psi = x$ in the above equation (\ref{30}), we get
\begin{equation}
\mathcal N = \frac{2}{\pi} \int_{0}^{\sqrt R} \frac{dx}{1-x^2}
\sqrt{\mathcal E^2(1-x^2) -m^2 x^2(1-x^2) - \mathcal J^2},
\label{31}
\end{equation}
where $R = \frac{-(m^2+ \mathcal E^2)+ \sqrt{(m^2+\mathcal E^2)^2
- 4 m^2 (\mathcal E^2 -\mathcal J^2) }}{-2 m^2}$,\\
and
\begin{eqnarray}
\frac{\partial \mathcal N}{\partial m} &=& -\frac{2m}{\pi}
\int_{0}^{\sqrt R}\frac{x^2}{\sqrt{\mathcal E^2(1-x^2) -m^2
x^2(1-x^2) - \mathcal J^2}}dx. \nonumber \\
%\label {32}
%\end{equation}
%This gives us elliptical integral which is
%\begin{equation}
%\frac{\partial \mathcal N}{\partial m}
&=& \frac{\sqrt{2a_+}}{\pi
m} \left[\mathbb {E}\Big(\frac{a_-}{a_+}\Big) - \mathbb
{K}\Big(\frac{a_-}{a_+}\Big)\right], \label{33}
\end{eqnarray}
where $a_\pm = (m^2+\mathcal E^2) \pm \sqrt{(m^2+\mathcal E^2)^2 -
4 m^2(\mathcal E^2- \mathcal J^2)}$ and a condition of
$(m^2-\mathcal E^2)^2 + 4m^2 \mathcal J^2 > 0$, which give an
upper bound to the $\mathcal N$. $\mathbb{E}$ and $\mathbb{K}$ are
the usual Elliptic integral of first and second kind respectively.
Now expanding equation (\ref{33}) in small $\mathcal E$ and
$\mathcal J$ for small $\mathcal N$, we get
\begin{eqnarray}
\frac{\partial \mathcal N}{\partial m} &=& \left[\frac{\mathcal
J^2} {2m^2} - \frac{15\mathcal J^4}{16 m^4} + \mathcal O[\mathcal
J^6]\right] + \left[-\frac{1}{2 m^2} + \frac{9 \mathcal J^2}
{8m^4} - \frac{525\mathcal J^4}{128 m^6} + \mathcal O[\mathcal
J^6]\right] \mathcal E^2 \cr && \cr &+& \left[-\frac{3}{16
m^4}+\frac{225\mathcal J^2} {128m^6} + \frac{11025\mathcal
J^4}{1024 m^8} + \mathcal O[\mathcal J^6]\right] \mathcal E^4 +
\mathcal O[\mathcal E^6]. \label{34}
\end{eqnarray}
Integrating the above equation (\ref{34}) with respect to $m$ and
reversing the series, we get
\begin{equation}
\mathcal E = \sqrt{2m \mathcal L}~~ K_3(\mathcal J)\left[ 1 -
K_4(\mathcal J) \frac{\mathcal L}{8m}+ \mathcal O[\mathcal
L^2]\right], \label{35}
\end{equation}
\begin{eqnarray}
{\rm where} ~~~\mathcal L &=& \mathcal N + \frac{\mathcal J^2}{2
m}-\frac{5\mathcal J^4}{16 m^3}+ \mathcal O[\mathcal J^6] \cr &&
\cr K_3(\mathcal J) &=& \left[1 - \frac{3\mathcal J^2}{4
m^2}+\frac{105\mathcal J^4}{64 m^4}+ \mathcal O[\mathcal
J^6]\right]^{-1/2} \cr && \cr {\rm and}~~~ K_4(\mathcal J) &=&
\left[1 - \frac{45\mathcal J^2}{8 m^2}+\frac{1575\mathcal J^4}{64
m^4}+ \mathcal O[\mathcal J^6]\right] K^4_3(\mathcal J).
\label{36}
\end{eqnarray}
The equation (\ref{35}) gives the short string oscillation energy
in $\mathbb {R} \times S^3$ with an extra angular momentum. This
is the same result of \cite{Beccaria:2010zn}, except we have
$\mathcal J$ dependent factor with every term in the expression.
The one loop correction to the energy of this type of pulsating
string can be calculated following \cite{Beccaria:2010zn} and
appropriate correspondence to those kind of configurations can be
made.
\section{Pulsating and rotating string with two spins
in $AdS_5 \times S^1$} In this section we wish to study a class of
long pulsating strings (in the large energy limit) which is
pulsating as well as rotating in the $AdS_5$ space. It has two
spin in $AdS$ which are mutually perpendicular and has an angular
momentum along the $S^1 \subset S^5$. We get the metric for $AdS_5
\times S^1$ background by putting $\theta = \xi_i = 0$ and $\phi =
\frac{\pi}{4}$ in the equation (\ref{A})
\begin{eqnarray}
ds^2 &=& -\cosh^2\rho dt^2 + d\rho^2 + \frac{1}{2}\sinh^2\rho
(d\phi^2_1+d\phi^2_2) + d\psi^2. \label{37}
\end{eqnarray}
Now the Polyakov action for the string in the above background is
given by
\begin{equation}
I = \frac{\sqrt\lambda}{4\pi}\int d\tau d\sigma\Big[-\cosh^2\rho
(\dot{t}^2 - {t^{\prime}}^2) + \dot{\rho}^2 - {\rho^{\prime}}^2 +
\frac{1}{2} \sinh^2\rho(\dot{\phi}^2_1+\dot{\phi}^2_2 -
{\phi^{\prime}}^2_1-{\phi^{\prime}}^2_2) + \dot{\psi}^2 -
{\psi^{\prime}}^2 \Big]. \label{38}
\end{equation}
We take the following ansatz for studying the pulsating string in
the above background
\begin{eqnarray}
&&t = t(\tau), ~~ \rho = \rho(\tau), ~~ \phi_1 = \phi(\tau)+ m
\sigma, ~~~ \phi_2 = \phi(\tau)- m \sigma,~~ \psi = \psi(\tau).
\label{39}
\end{eqnarray}
The equations of motion for $t$, $\rho$ and $\phi_1$ and $\phi_2$
are
\begin{eqnarray}
&&2\cosh\rho \sinh\rho \dot{\rho}\dot t + \cosh^2\rho \ddot t =
0,\cr & \cr &&\ddot\rho + \sinh\rho \cosh\rho (m^2 + \dot{t}^2 -
\dot\phi^2) = 0 , \cr & \cr &&2\cosh\rho \sinh\rho
\dot{\rho}\dot\phi + \sinh^2\rho \ddot\phi = 0. \label{40}
\end{eqnarray}
The Virasoro constraint gives us
\begin{equation}
\dot\rho^2 + \dot\psi^2 + m^2\sinh^2\rho + \dot\phi^2 \sinh^2\rho
- \dot t^2 \cosh^2\rho = 0. \label{41}
\end{equation}
Now the conserved quantities are as follows
\begin{equation}
\mathcal E = \dot t \cosh^2\rho, ~~~~~~ \mathcal S_1 =\mathcal S_2
=\mathcal S = \frac{1}{2} \dot\phi \sinh^2\rho ~~~~~~ \mathcal J =
\dot\psi, \label{42}
\end{equation}
where $\mathcal S = \frac{S}{\lambda}$. Putting the above equation
(\ref{42}) in the equation (\ref{41}), we get
\begin{equation}
\dot\rho^2 = \frac{\mathcal E^2}{\cosh^2\rho} -\frac{4\mathcal
S^2}{\sinh^2\rho} - m^2\sinh^2\rho -\mathcal J^2. \label{43}
\end{equation}
If we consider the above equation as the equation of motion of a
test particle moving in a potential, then the particle experiences
infinite potential at both zero and infinity having a minimum in
between. So the radial coordinate oscillates in between a minimum
($\rho_{min}$) and maximum value ($\rho_{max}$). Now the
oscillation number
\begin{equation}
\mathcal N = \frac{1}{2\pi} \oint d\rho \dot\rho = \frac{1}{\pi}
\int_{\rho_{min}}^{\rho_{max}} d\rho \sqrt{\frac{\mathcal
E^2}{\cosh^2\rho} -\frac{4\mathcal S^2}{\sinh^2\rho} -
m^2\sinh^2\rho -\mathcal J^2 }.\label{44}
\end{equation}
Putting $\sinh\rho = x$ in the above equation (\ref{44}), we get
\begin{equation}
\mathcal N = \frac{1}{\pi} \int_{\sqrt R_1}^{\sqrt R_2}
\frac{dx}{1+x^2} \sqrt{\mathcal E^2 - \frac{4 \mathcal
S^2(1+x^2)}{x^2} -m^2 x^2(1+x^2) - \mathcal J^2(1+x^2)},
\label{45}
\end{equation}
where $R_1$ and $R_2$ are two positive roots of the cubic
polyomial
\begin{equation}
f(z) = m^2 z^3 + (m^2+ \mathcal J^2) z^2 + (4\mathcal S^2 +
\mathcal J^2 - \mathcal E^2) z + 4\mathcal S^2, ~~~~~~~~~ z \equiv
x^2, \label{B}
\end{equation}
with the condition $\mathcal E^2 \geq 4\mathcal S^2 +\mathcal
J^2$. Now taking the partial derivative of the equation
({\ref{45}) with respect to $m$ we get
\begin{equation}
\frac{\partial \mathcal N}{\partial m} = -\frac{m}{\pi}
\int_{\sqrt R_1}^{\sqrt R_2}\frac{x^2}{\sqrt{\mathcal E^2 -
\frac{4 \mathcal S^2(1+x^2)}{x^2} - m^2 x^2(1+x^2) - \mathcal
J^2(1+x^2)}}dx. \label {46}
\end{equation}
This can be written, by using the usual elliptical integral of
first and second kind, as
\begin{equation}
\frac{\partial \mathcal N}{\partial m} = \frac{1}{\pi}
\frac{1}{\sqrt{R_2-R_3}} \left[(R_3-R_2)\mathbb
{E}\Big(\frac{R_2-R_1}{R_2-R_3}\Big) - R_3 \mathbb
{K}\Big(\frac{R_2-R_1}{R_2-R_3}\Big)\right], \label{47}
\end{equation}
For large $\mathcal E$, but small $\mathcal S$ and $\mathcal J$,
we get the roots as
\begin{eqnarray}
&&R_{2,3} = \pm \frac{\mathcal E}{m} -\frac{1}{2} -\frac{\mathcal
J^2}{2m^2} \pm \frac{m^2-2\mathcal J^2- 16 \mathcal S^2}{8m
\mathcal E} + \mathcal O[\mathcal E^{-2}],\cr & \cr &&R_1 =
\frac{4\mathcal S^2}{\mathcal E^2} + \mathcal O[\mathcal E^{-4}].
\label{48}
\end{eqnarray}
Now expanding equation(\ref{48}) for large $\mathcal E$ we get
\begin{eqnarray}
\frac{\partial \mathcal N}{\partial m} &&= c_1 m^{-1/2}\mathcal
E^{1/2} + c_2 m^{1/2} \mathcal E^{-1/2} \Big(1+ \frac{\mathcal
J^2}{m^2} \Big) \cr && \cr &&+ m^{3/2} \mathcal E^{-3/2}\Big(c_3 +
c_4\frac{\mathcal J^2}{m^2} + c_5 \frac{\mathcal J^4}{m^4} +
c_6\frac{\mathcal S^2}{m^2} \Big) + \mathcal O[\mathcal E^{-5/2}],
\label{49}
\end{eqnarray}
where $c_1 = 4\frac{\sqrt{2\pi}}{\lceil[-\frac{1}{4}]^2} -
\frac{\mathbb{E}(-1)}{\pi} = -0.19069$, ~~~~~ $c_2 =
\frac{\sqrt{2\pi}}{\lceil[-\frac{1}{4}]^2} = 0.104328$,\\
$c_3 = -\frac{3 \sqrt{\pi}}{16 \sqrt 2 \lceil[\frac{1}{4}]^2} =
-0.0178772$, ~~~~~ $c_4 = \frac{\sqrt{2\pi}}{9
\lceil[-\frac{3}{4}]^2} = 0.0119181$,\\ $c_5 = -
\frac{\sqrt{\pi}}{16 \sqrt 2 \lceil[\frac{1}{4}]^2} = -0.00595906$
~and~ $c_6 = \frac{\sqrt{2\pi}} {\lceil[\frac{1}{4}]^2} =
0.19069$. Integrating the above equation (\ref{49}), we get
\begin{eqnarray}
\mathcal N = \mathcal N_0 &+& 2c_1 m^{1/2}\mathcal E^{1/2} +
\frac{2}{3} c_2 m^{3/2} \mathcal E^{-1/2} \Big(1 - 3
\frac{\mathcal J^2}{m^2}\Big) \cr && \cr &+& m^{5/2} \mathcal
E^{-3/2}\Big(\frac{2}{5} c_3 + 2c_4 \frac{\mathcal J^2}{m^2} -
\frac{2}{3} c_5 \frac{\mathcal J^4}{m^4} + 2 c_6 \frac{\mathcal
S^2}{m^2}\Big) + \mathcal O[\mathcal E^{-5/2}]. \label{50}
\end{eqnarray}
The integration constant $\mathcal N_0$ can be determined from the
integral (\ref{45}) for $m = 0$
\begin{equation}
\mathcal N_0 = \frac{1}{\pi}
\int_{r_1}^{r_2}\frac{dx}{1+x^2}\sqrt{\mathcal E^2
-\frac{4\mathcal S^2(1+x^2)}{x^2} -\mathcal J^2 (1+x^2)},
\label{51}
\end{equation}
where $r_{1,2}^2 = \frac{\mathcal E^2 - 4\mathcal S^2 -\mathcal
J^2\mp \sqrt{(\mathcal E^2 - 4\mathcal S^2 -\mathcal J^2)^2 - 16
\mathcal S^2 \mathcal J^2}}{2\mathcal J^2}$. Changing the variable
we get
\begin{equation}
\mathcal N_0 = \frac{r_1}{\pi} \int_{1}^{r_2/r_1}
\frac{dx}{1+r^2_1 x^2}\sqrt{r^2_1\mathcal
J^2(1-x^2)+\frac{4\mathcal S^2}{r^2_1}(1-\frac{1}{x^2})}
 = \frac{1}{2} (\mathcal
E-2\mathcal S-\mathcal J). \label{52}
\end{equation}
%
%\begin{equation}
%\mathcal N = \frac{1}{2} (\mathcal E - 2\mathcal S - \mathcal J) +
%2c_1 m^{1/2}\mathcal E^{1/2} + \frac{2}{3} c_2 m^{3/2} \mathcal
%E^{-1/2} + \frac{2}{5} c_3 m^{5/2} \mathcal E^{-3/2} + \mathcal
%O[\mathcal E^{-5/2}]. \label{53}
%\end{equation}
%
Now putting (\ref{52})in (\ref{50}) and reversing the series we
get
\begin{equation}
\mathcal E = 2L + a_1 m^{1/2} L^{1/2} + a_2 m - A(\mathcal J) a_3
m^{3/2} L^{-1/2} + B(\mathcal J, \mathcal S) a_4 m^{5/2} L^{-3/2}
+\mathcal O [L^{-5/2}], \label{54}
\end{equation}
where ~~~$L = \mathcal N + \mathcal S + \frac{\mathcal
J}{2},~~~~~~ B(\mathcal J, \mathcal S) = 1 - 3.48 \frac{\mathcal
J^2}{m^2} - 0.35 \frac{\mathcal J^4}{m^4} -  34.05 \frac{\mathcal S^2}{m^2}$\\
and $ a_4 = 0.00791996 $. In the above (\ref{54}) series $\mathcal
O[L^{-3/2}]$ term on wards the coefficients will depend on even
power of angular momentum $\mathcal J$ and spin $\mathcal S$
explicitly along with $L$.
%In each step there will be additon of
%one $\mathcal J$ and one $\mathcal S$ terms to the $\mathcal J$
%$\mathcal S$ dependent terms similar to the (\ref{20})
The above
equation (\ref{54}) is the energy expression for the long strings
with small $\mathcal J$ and $\mathcal S$ in $AdS_5 \times S^2$.
This is the same expression as in \cite{Minahan:2002rc} with
$\mathcal S =\mathcal J = 0$ . In the above equation if we put
$\mathcal S = 0$, we get back to equation (\ref{20}).
\section{Conclusion}
In this paper, we have studied some examples of pulsating strings
with extra angular momentum along various subspace of the $AdS_5
\times S^5$. First we have found the energy for the short and long
strings with small $\mathcal J$ and for the short string with
large $\mathcal J$ in $AdS_3 \times S^1$. Then we have found the
short string solution with small angular momentum in the $\mathbb
{R} \times S^3$ background. Further we have shown the long spin
energy behavior in the small $\mathcal S$ and $\mathcal J$ in
$AdS_5 \times S^1$. An interesting study will be to find out the
nature of gauge theory operators in the dual CFT. Similar studies
have been performed earlier in, e.g.
\cite{Park:2005kt}\cite{Russo:2002sr}\cite{Beisert:2003ea}\cite{Ferretti:2004ba}.
The operator for a rotating string in $AdS_5 \times S^5$ with
charge $(S,J)$ has been suggested to be $Tr~\mathcal D^S Z^J$
\cite{Beisert:2003ea} where $\mathcal D$ is the complex
combination of covariant derivative and $Z$ is the complex scalar.
Similarly some generic operator, dual to spinning and rotating
string has been shown in \cite{Russo:2002sr}, where the exact
operator is a linear combination of operators containing $S$
insertions of $\mathcal D_i$ into $Tr~Z^J$ for specific $(S,J)$
charge conditions. Furthermore, in \cite{Park:2005kt} for a
rotating and pulsating string in $AdS_5$ with two equal spins in
two orthogonal directions, the dual operator has been suggested to
be of the form $Tr ~\mathcal D^S_X \mathcal D^S_Y$ (where
$\mathcal D_X = \mathcal D_1 + i \mathcal D_2$ and $\mathcal D_Y =
\mathcal D_3 + i \mathcal D_4$) which is self dual components of
gauge field strength \cite{Ferretti:2004ba}. We expect the
solution presented in the chapter-4 may be dual to the gauge field
operator of the generic form $Tr ~\mathcal D^S_X \mathcal D^S_Y
Z^J$. The anomalous dimensions and the dual gauge theory operators
for the string solutions presented here are under construction.

\vskip .2in \noindent {\bf Acknowledgements:} We would like to
thank anonymous referee for constructive suggestions. KLP would
like to thank the Abdus Salam I.C.T.P, Trieste for hospitality
under Associate Scheme, where a part of this work was completed.

\end{document}